\documentclass{elsart}

\usepackage{epsfig}

\def\lsim{\lower.5ex\hbox{$\; \buildrel < \over \sim \;$}}
\def\gsim{\lower.5ex\hbox{$\; \buildrel > \over \sim \;$}}

\newcommand{\mdot}{\mbox{$\dot{M}$}}

\usepackage{amssymb}

\def\url#1{{\ttfamily\def\/{/\discretionary{}{}{}}#1}}

\begin{document}

\begin{frontmatter}
\title{Two temperature viscous accretion
flows around rotating black holes: Description of under-fed systems to ultra-luminous X-ray sources}
\author{S. R. Rajesh\thanksref{email1}},
\author{Banibrata Mukhopadhyay\thanksref{email2}}
\address{Department of Physics, Indian Institute of Science, Bangalore 560012, India}
\thanks[email1]{rajesh@physics.iisc.ernet.in}
\thanks[email2]{bm@physics.iisc.ernet.in}

\begin{abstract}
We discuss two temperature accretion disk flows around rotating
black holes.
As we know that to explain observed hard X-rays the choice of Keplerian angular
momentum profile is not unique, we consider the sub-Keplerian regime of
the disk. Without any strict knowledge of the magnetic field structure, we assume
the cooling mechanism is dominated by bremsstrahlung process.
We show that in a range of Shakura-Sunyaev viscosity parameter
$0.2\gsim\alpha\gsim0.0005$, flow behavior varies widely, particularly by means of
the size of disk, efficiency of cooling and corresponding temperatures of
ions and electrons. We also show that the disk around a rotating black hole
is hotter compared to that around a Schwarzschild black hole, rendering
a larger difference between ion and electron temperatures in the former case.
With all the theoretical solutions in hand, finally we reproduce the observed luminosities ($L$)
of two extreme cases --- the under-fed AGNs and
quasars (e.g. Sgr~$A^*$) with $L\gsim 10^{33}$ erg/sec to ultra-luminous X-ray sources
with $L\sim 10^{41}$ erg/sec,
at different combinations of mass accretion rate, ratio of
specific heats, Shakura-Sunyaev viscosity parameter and Kerr parameter, and
conclude that Sgr~$A^*$ may be an intermediate spinning black hole.
\end{abstract}

\end{frontmatter}

\section{Introduction}

It was shown by Shapiro, Lightman \& Eardley (1976) that to explain observed 
hard X-rays e.g. from Cyg~X-1 the simple Keplerian accretion disk 
(Shakura \& Sunyaev 1973; Novikov \& Thorne 1973) is inappropriate.
Hence they proposed a two temperature accretion disk with electron and ion
temperatures respectively $\sim 10^9$K and $\sim 5\times 10^{11}$K.
Indeed Eggum et al. (1985) showed that the cool disk flow with a constant 
viscosity parameter $\alpha$ and Keplerian angular
momentum profile, which is optically thick and geometrically thin, collapses.

Paczy\'nski \& Wiita (1980) proposed a two component accretion disk 
which is geometrically thick in the optically thick
limit. Subsequently, Rees et al. (1982) introduced hot ion torus model
in the optically thin limit and Muchotrzeb \& Paczy\'nski (1982) initiated
the model of transonic accretion in the sub-Keplerian regime. 
The idea was further pronounced by Chakrabarti (1996) emphasizing the formation
of shock in flows and Mukhopadhyay (2003),
Mukhopadhyay \& Ghosh (2003; hereafter MG03) showing importance of rotation of
the central compact object, in full general relativistic as well as
pseudo-general relativistic (Mukhopadhyay 2002) frameworks.
In the regime of inefficient cooling Narayan \& Yi (1995) 
introduced a two temperature disk model based on self-similar solutions. 
Abramowicz et al. (1988), on the other hand, proposed a height-integrated disk flow 
at super-Eddington accretion rate when the diffusion time scale is longer than
the viscous time scale rendering a high optical depth of the flow.
Other works which helped in developing the modern accretion disk theory
may be named as those by, e.g., Begelman (1978), Liang \& Thompson (1980), 
Eggum, Coroniti \& Katz (1988). 

The two temperature model by Shapiro, Lightman \& Eardley (1976) was significantly
hotter than single temperature model of Shakura \& Sunyaev (1973).
The hot gas which is optically thin, however thermally unstable, 
is considered to be cooled down through the
bremsstrahlung and inverse-Compton
processes and therefore could explain various states of Cyg~X-1 (Melia \& Misra 1993).
On the other hand, the ``ion torus" model by Rees et al. (1982) could explain
AGNs at a low mass accretion rate. 

However, none of the models attempt to describe the two temperature flow around 
a rotating black hole and to understand any effect of nonzero specific angular
momentum of the black hole. This is particularly important in understanding the
inner disk properties. 
Moreover, there is no attempt, to best of our knowledge,
to understand the variation of profiles of ion and electron temperatures 
in a wide range of $\alpha$ proposed by Shakura-Sunyaev (1973).
For example, while Mandal \& Chakrabarti (2005) modeled a two temperature flow,
they restricted with static black holes without uncovering specific values of $\alpha$
required for realistic solutions. 
Their particular emphasis
was how the shock helps in fulfilling the radiative processes, and thus they
straight away supplied the compression ratio of the flow at the shock location as an input without
solving the set of hydrodynamic equations self-consistently. Recently, following the same approach,
Das \& Chattopadhyay (2008) computed mass loss from viscous accretion disk in presence of cooling.

Very recently, Ghosh \& Mukhopadhyay (2009) and Ghosh et al. (2010) described
ultra-luminous sources to under-luminous AGNs based on 2.5-dimensional
self-similar solutions of accretion disks. However, they could not include
the information of angular momentum of the black hole to the solutions due to very nature of
the solution procedure (strict choice of self-similarity). Moreover, in their
self-similar approach they could not explore two temperature structure of accretion disks,
which eventually is an important factor in determining efficiency of cooling and
then luminosity. 

In the present paper, we plan to model a two temperature relativistic 
transonic sub-Keplerian accretion flow. 
We consider the set of hydrodynamic equations describing disk flows
along with the components of cooling processes, particularly bremsstralung radiation.
This is now easier to work out by using the pseudo-Newtonian potential given by
Mukhopadhyay (2002), which is being considered to model accretion flows around
rotating black holes for sometimes (e.g. Chan, Psaltis \& Feryal 2005;
Lipunov \& Gorbovskoy 2007; Stuchlik \& Kov\'ar 2008; Shafee, Narayan \& McClintock 2008;
Benson \& Babul 2009).
However, we neither restrict, unlike Narayan \& Yi (1995), to the radiatively inefficient advection 
dominated regime of self-similarity nor, unlike Mandal \& Chakrabarti (2005),
to a particular aspect of dynamics, e.g. possible formation of shock.
We then focus on two extreme regimes of the disk flow --- the under-luminous AGNs 
and quasars (e.g. Sgr~$A^{*}$) and 
ultra-luminous X-ray (ULX) sources (e.g. SS433), to implement our model
in explaining their luminosities and extracting fundamental properties, e.g. spin.

In the next section, we present the model equations and outline
the procedure to solve them. Subsequently, we discuss
the two temperature accretion disk solutions around stellar mass
and supermassive black holes, respectively for super-Eddington and 
sub-Eddington accretion rates, in \S3.
Section 4 compares the disk flow around a rotating black hole with the flow of
similar parameters around a static black hole. Finally we discuss 
the implications of results with a summary in \S5.

\section{Model equations describing the system}

To compute hydrodynamic variables we strictly follow Chakrabarti (1996),
Mukhopadhyay (2003) and MG03,
but with the gravitational force [as of Mukhopadhyay (2003)] for a black hole given 
by Mukhopadhyay (2002). Therefore, without
discussing the insight, below we straight away recall the basic
equations.  Note that throughout we express all the
variables having their usual meaning in dimensionless units as described in MG03, 
unless stated otherwise. Hence
\begin{eqnarray}
\mdot\,=\,-4\pi x \Sigma \vartheta,
\label{mass}
\end{eqnarray}
where the surface density
\begin{eqnarray}
\Sigma \,= \,I_n\, \rho_{eq} h(x),\,\,
I_n \,= \,(2^{n} n!)^{2}/(2n+1)!\,\, {\rm (Matsumoto\hskip0.2cm et\hskip0.2cm al.\hskip0.2cm 1984)},
\label{matsu}
\end{eqnarray}
when $\rho_{eq}$ is the density at the equatorial plane and the half-thickness is given by
\begin{eqnarray}
h(x) \,= \,c_s x^{1/2} F^{-1/2}.
\label{thik}
\end{eqnarray}
\begin{eqnarray}
\vartheta \frac{d\vartheta}{dx} \,+ \,\frac{1}{\rho} \frac{dP}{dx} \,- \,\frac{\lambda^{2}}{x^3} \,+ \,F \, = \, 0,
\label{rad}
\end{eqnarray}
when we consider the adiabatic equation of state $P=K\rho^\gamma$ 
with $\gamma$ being the ratio of specific heats ranging from
$4/3$ to $5/3$,
and following Mukhopadhyay (2002)
\begin{eqnarray}
F=\frac{(x^2 - 2a\sqrt{x} +a^2)^2}{x^3[\sqrt{x}(x - 2) +a]^2},
\end{eqnarray}
where $a$ is the specific angular momentum (Kerr parameter) of the black hole.
We also define 
\begin{eqnarray}
\nonumber
\beta=\frac{{\rm gas\hskip0.2cm pressure\hskip0.2cm} P_{gas}}
{{\rm total\hskip0.2cm pressure\hskip0.2cm} P}\,=\frac{6\gamma-8}{3(\gamma-1)}
\,\,\,\, {\rm (e.g.\hskip0.2cm Ghosh\hskip0.2cm \&\hskip0.2cm Mukhopadhyay\hskip0.2cm 2009)},\\
\end{eqnarray}
where $P_{gas}=P_i\hskip0.2cm({\rm ion\hskip0.2cm pressure})
+P_e\hskip0.2cm({\rm electron\hskip0.2cm pressure})$, such that
\begin{eqnarray}
P \,=\,\frac{\rho}{\beta\,c^2} \left(\frac{kT_{i}}{\mu_{i} m_{i}} \,+ 
\,\frac{kT_{e}}{\mu_{e} m_{i}}\right) \,= \,\rho c^{2}_{s}.
\label{ptot}
\end{eqnarray}
Here $T_{i}$, $T_{e}$ are respectively the ion and electron temperatures in Kelvin, 
$m_i$ the mass of proton in gm, $\mu_i$ and $\mu_e$
respectively the corresponding effective molecular weight, $c$ the speed
of light, $k$ the Boltzmann constant.  
\begin{eqnarray}
\vartheta \frac{d \lambda}{dx} \,= \,\frac{1}{\Sigma x} \frac{d}{dx}\left(x^2 |W_{x \phi}|
\right), 
\label{az}
\end{eqnarray}
where 
\begin{eqnarray}
W_{x \phi}  \,= \,- \alpha  \left(I_{n+1} P_{eq} \,+ \,I_n \vartheta^2 \rho_{eq} \right)h(x)\,\,\,
{\rm (Chakrabarti\,\,1996,\,\,MG03)}, 
\end{eqnarray}
when $P_{eq}$ is the pressure 
at equatorial plane. Note that we will assume $P_{eq}\sim P$ and
$\rho_{eq}\sim\rho$ in obtaining solutions. 
Here we assume a random isotropic magnetic field in the disk governing $\alpha$.

\begin{eqnarray}
\frac{\vartheta h(x)}{\Gamma_{3} - 1} \left(\frac{dP}{dx} \,- \,\Gamma_{1} \frac{P}{\rho} \frac{d \rho}{dx}\right) \,= \,f\,Q^{+},
\label{eni}
\end{eqnarray}
where 
\begin{eqnarray}
\nonumber
\Gamma_{3} &= &1 \,+ \,\frac{\Gamma_{1} \,- \,\beta}{4 \,- \,3 \beta},\\
\nonumber
\Gamma_{1} &= &\,\beta \,+ \,\frac{(4 \,- \,3 \beta)^{2}(\gamma \,- \,1)}{\beta+\,12(\gamma \,- \,1)(1 \,- \,\beta)}\\
Q^+ &= & \alpha (I_{n+1} P \,+ \,I_n \vartheta^2 \rho )h(x) \frac {d \lambda}{dx}\,\,\, {\rm (Chakrabarti\,\,1996,\,\,
MG03),}
\label{qvis}
\label{gam}
\end{eqnarray}
when $Q^+$ is the heat generated by viscous dissipation and $f$ the parameter determining
the fraction of energy advected into the black hole.

Now combining above equations appropriately we obtain (MG03)
\begin{eqnarray}
\frac{d\vartheta}{dx} \,= \,\frac{f_1(x,\vartheta,c_{s},\lambda)}{f_2(\vartheta,c_{s})},
\label{dvdx}
\end{eqnarray}
where $f_1(x,\vartheta,c_{s},\lambda)$, $f_2(\vartheta,c_{s})$ and the solution
procedure in detail are given in MG03 (see also Chakrabarti\,\,1996). 

After obtaining the hydrodynamic solutions we plug them into the electron heating 
equation given by  
\begin{eqnarray}
\frac{\vartheta h(x)}{\Gamma_{3} - 1} \left(\frac{dP_{e}}{dx} \,- \,\Gamma_{1} \frac{P_{e}}{\rho} \frac{d \rho}{dx}\right) \,= \,Q_{ie} \,-\,Q^{-},
\label{ene}
\end{eqnarray}
where $Q_{ie}$ is
the Coulomb coupling (Bisnovatyi-Kogan \& Lovelace 2000) given in dimensionful
unit as
\begin{eqnarray}
q_{ie} \, = \,\frac{8 (2 \pi)^{1/2}e^4 n_i n_e}{m_i m_e}\left(\frac{T_e}{m_e} \,+ \, \frac{T_i}{m_i}\right)^{-3/2} \ln (\Lambda) \ \left(T_i \, - \,T_e \right)\,\,{\rm erg/cm^3/sec},
\label{qie}
\end{eqnarray}
$n_i$ and $n_e$ denote the number densities of ion and electron respectively, 
$e$ the charge of an electron,
ln($\Lambda$) the Coulomb logarithm. Then
$Q^-$ may be the heat radiated away by the
bremsstrahlung, synchrotron processes and inverse Comptonization
of soft photons. For the present purpose without a proper knowledge of
the magnetic field structure in disk, we assume any radiative loss is due to
the bremsstrahlung effect
given in dimensionful form as (see Narayan \& Yi 1995;
Mandal \& Chakrabarti 2005 for detailed description, what we do not repeat here)
\begin{eqnarray}
q_{br} &= &1.4 \times 10^{-27} \ n_e\,n_i T_e^{1/2}\,(1+4.4\times 10^{-10} T_e)\,
\,\,{\rm erg/cm^3/sec}.
\label{qvari}
\end{eqnarray}
However, in future we plan to include other components of cooling processes,
e.g. synchrotron, inverse-Compton effect to make the model more general (Sinha, Rajesh
\& Mukhopadhyay  2009;
Rajesh \& Mukhopadhyay 2009).
Important point to note is that in principle one should solve all the five 
conservation equations simultaneously to obtain the solution 
(e.g. Sinha, Rajesh \& Mukhopadhyay 2009). 
However, for the present purpose we simply feed the solutions of the hydrodynamic set 
into the electron energy conservation equation and obtain the electron temperature.
We take the present analysis as an exploratory study. If we find that in a
wide range of $\alpha$, as well as in the range of Kerr parameter, 
the solutions vary noticeably bringing important insights, then we would like to analyze
the detailed general solutions by solving the five coupled equations (1), (4), (8), (10), (13)
(Rajesh \& Mukhopadhyay 2009) simultaneously.

As the flow is considered to be
of two temperatures, at the sonic radius $x_c$ the electron temperature $T_{ec}$ needs to be
specified, along with corresponding specific angular momentum $\lambda_c$ of the disk. 
Note that one has to self-consistently adjust the set of values $x_c, \lambda_c, T_{ec}$ 
to obtain solutions connecting
outer boundary and black hole event horizon through $x_c$.
Moreover, $\gamma$ of the flow to be specified according to $\mdot$ --- higher the
$\mdot$, higher the rate of supply of radiation, lower the $\gamma$ is and vice versa.  
The outer boundary corresponds to the radius $x_o$ where $\lambda/\lambda_K=1$.
($\lambda_K$ being the specific angular momentum of the flow when the centrifugal force is 
same as gravitational force).  Throughout in the text, we quote $x_o$ as the Keplerian to
sub-Keplerian transition radius, while we essentially restrict to the
inner sub-Keplerian disk.


Below we discuss the solutions particularly for: (1) super-Eddington accretion around stellar
mass black holes, (2) sub-Eddington accretion around supermassive black holes.

\section{Super- and sub-Eddington accretion around black holes}

\subsection{Super-Eddington accretion around stellar mass black holes}

The ``radiation trapped'' accretion disk can be attributed to
the radiatively driven outflows or jets.
This presumably occurs when the accretion rate is super-Eddington
(Lovelace et al. 1994, Begelman et al. 2006, Fabbiano 2004, Ghosh \&
Mukhopadhyay 2009),
as seen in ULX sources such as SS433 (with luminosity
$\gsim 10^{40}$ erg/sec or so; Fabrika 2004). In order to describe such sources, the
models described below are the meaningful candidates.

We choose the accretion rate $\mdot=10$ (throughout, the accretion rate is expressed in 
Eddington units) and mass of the black hole $M=10$ (throughout, the mass of black hole
is expressed in units of solar mass) and obtain solutions for three values of 
$\alpha=0.2,0.01,0.0005$ around nonrotating (Schwarzschild) and rotating 
(Kerr with $a=0.998$) black holes. 


\subsubsection{Schwarzschild black holes}

\begin{figure}
\begin{center}
\rotatebox{0}{\epsfig{file=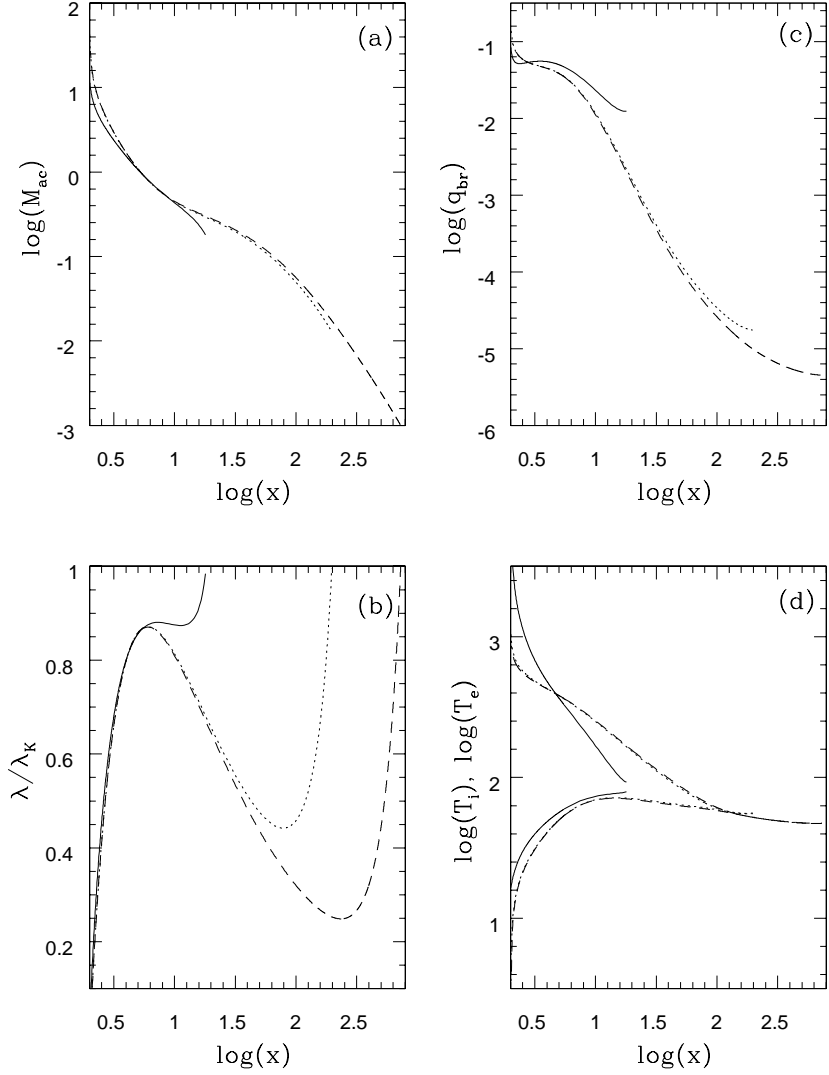,width=4.5in}}
\end{center}
\caption{ 
Variation of (a) Mach number, (b) ratio of disk angular momentum to corresponding
Keplerian angular momentum, (c) bremsstrahlung radiation in erg/sec, (d) electron
(lower set of curves) and ion (upper set of curves) temperatures in units of $10^9$K, as functions
of radial coordinate around a Schwarzschild black hole of $M=10$ and $\mdot=10$. 
The solid, dotted, dashed curves correspond to $\alpha=0.2,
0.01,0.0005$ respectively. The respective parameter sets are given in Table 1 in detail.
}\label{fig1}
\end{figure}

We first consider flows around static black holes where the Kerr parameter $a=0$.
Figure \ref{fig1} shows the behavior of flow variables as functions of
radial coordinate for $\alpha=0.2,0.01,0.0005$. Higher the $\alpha$, lower the 
residence time of the flow is, which results in a hotter flow with a larger $f$.
Therefore we choose the corresponding $f$ appropriately. Note that
Narayan \& Yi (1995) chose strongly advection dominated hot flows with $f=1$.
However, for the present purpose, we neither consider such an extreme case
nor prefer to restrict with advection dominated flows. We rather like
to consider flows of general advective paradigm.
The sets of input parameters
for the model cases described here are given in Table 1. Figure \ref{fig1}a
verifies that a lower $\alpha$ corresponds to a lower rate of energy momentum transfer
which implies a higher residence time of the flow in the sub-Keplerian regime.
As a result, the radial velocity shows a stronger centrifugal barrier for a
lower $\alpha$. As a consequence, Fig. \ref{fig1}b shows that $x_o$
recedes further out for a lower $\alpha$.
Figure \ref{fig1}c shows the bremsstralung cooling profile. Naturally the total
emission is higher for the flow with a lower $\alpha$ due to a higher residence time before
plunging into the black hole.
Therefore, the computed luminosity given in 
Table 1 is higher for a 
flow of lower $\alpha$, compared to that of a higher $\alpha$ with a similar parameter set.
Note that while a high $\alpha$ flow might generate higher dissipative energy compared 
to a low $\alpha$ flow, due to a lower residence time the former flow (with $\alpha\sim 0.2$) 
could not radiate out completely before plunging into the black hole and thus hotter. 
In this case, even the Keplerian-sub-Keplerian 
transition region is of two temperatures, unlike flows with lower $\alpha$.
Interestingly, in the vicinity of black hole the electron temperature goes down,
while the ion temperature goes up. 
This is because very strong advection close to the black hole 
renders a weaker ion-electron coupling. This in turn hinders the transfer
of energy from ions to electrons attributing ions to remain hot, while
electrons continue to cool down further by radiative processes.
Hence, the difference between ion and electron temperatures is higher for a higher
$\alpha$ obviously. 
However, as was shown by Yuan (2001) and Mandal \& Chakrabarti
(2005), while close to the black hole the advective heating to electrons 
becomes stronger, density of the flow increases sharply. 
This results in the increase of the cooling effects as well, which dominates
the advective heating. As electrons are decoupled from ions at
this regime, their temperature goes down.

\subsubsection{Kerr black holes}

\begin{figure}
\begin{center}
\rotatebox{0}{\epsfig{file=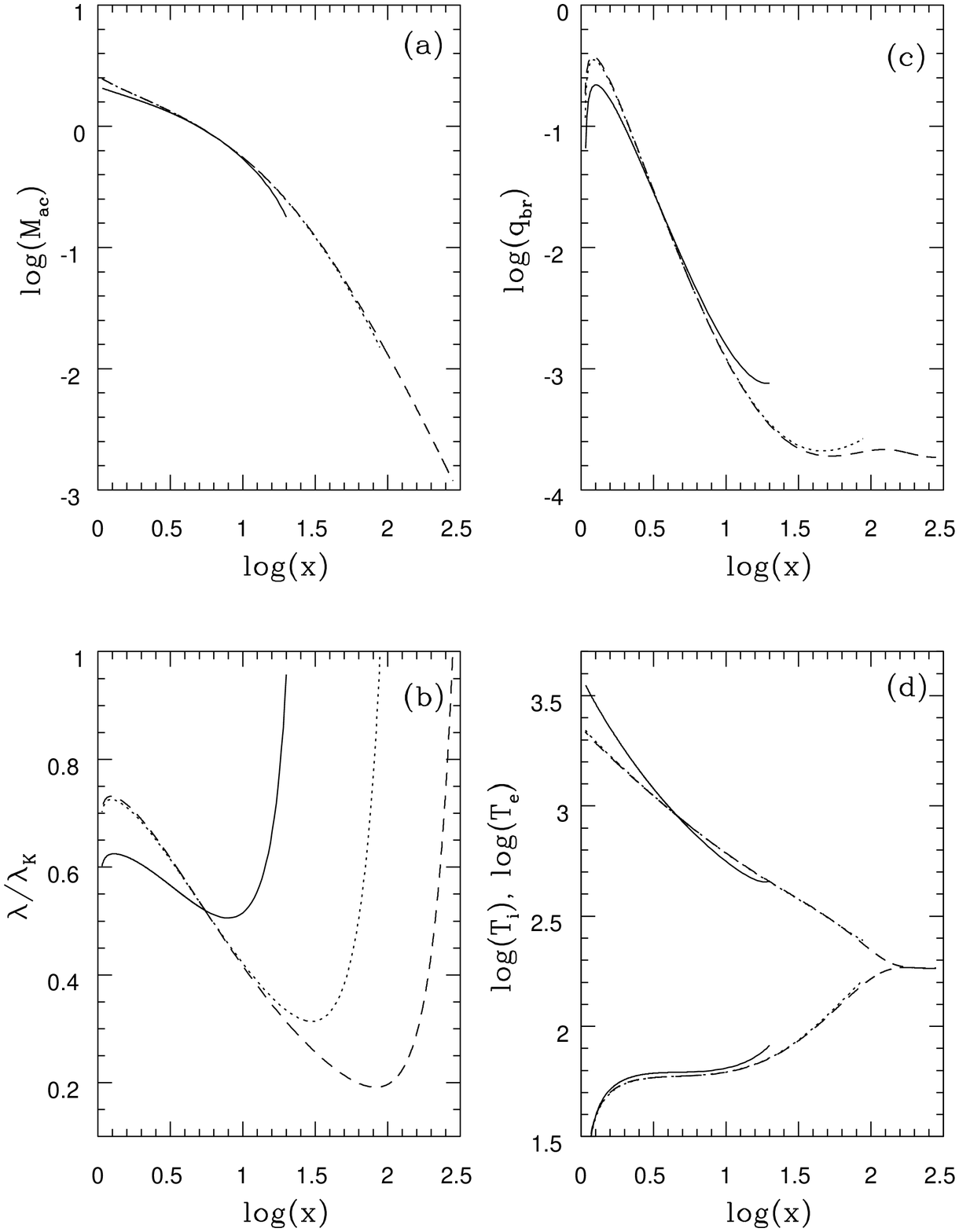,width=4.5in}}
\end{center}
\caption{ 
Same as Fig. \ref{fig1}, except for $a=0.998$
}\label{fig2}
\end{figure}

The specific angular momentum (Kerr parameter) for rotating black holes is chosen to be
$a=0.998$. As discussed earlier 
(Mukhopadhyay 2003), the angular momentum of the flow around a rotating black hole 
should be smaller compared to that around a static black hole. This reassembles
advancing $x_o$, which decreases the size of 
zone of present interest (see Fig. \ref{fig2}b). 
As the flow angular momentum decreases significantly (see Table 1),
the centrifugal barrier completely disappears as shown in Fig. \ref{fig2}a. Thus the infall time is
shorter compared to that around a static black hole attributing a smaller
residence time of the flow in an element of disk
hindering cooling processes to complete. Therefore, the entire disk is hotter compared to
that around a static black hole and, hence, even for $\alpha=0.01$ the transition zone
is of two temperatures, as shown in Fig. \ref{fig2}d.
Important point to note from Fig. \ref{fig2}c is that the decrease of residence time, close to the
black hole event horizon, hinders bremsstrahlung process severely which renders
a sharp downfall of bremsstrahlung radiation, compared to that around a static
black hole shown in Fig. \ref{fig1}c. 
As a result, the electron remains hotter compared to that around
a nonrotating black hole.

The disk luminosity around a rotating black hole decreases compared to that of a static black hole
when $\alpha$ is high. This is because, the relatively less radiative emission is
in the former flow when the disk size itself is smaller, as is the case
of high $\alpha$ compared to a flow of low $\alpha$. However, 
a higher $a$ corresponds to a smaller $\lambda$ which in turn decreases $\vartheta$ 
at a particular radius of the inner edge of disk, when the inner edge is stretched in, compared
to that around a Schwarzschild black hole. Therefore, at a low $\alpha$, this results in the 
increase of residence time and then radiative loss of the flow before plunging into the
black hole. Therefore, the luminosity increases compared to that from a disk around
a static black hole. See the luminosity column in Table 1.

\subsection{Sub-Eddington accretion around supermassive black holes}

The under-luminous AGNs and quasars (e.g. Sgr~$A^{*}$)
had been attempted to describe by Narayan \& Yi (1995)
where the flow is expected to be substantially sub-critical/sub-Eddington
with a very low luminosity ($\gsim 10^{33}$ erg/s).
Therefore the cases described below 
could be potential models in order to describe under-luminous
sources.

We choose, primarily, $\mdot=0.01$ and $M=10^7$ 
and obtain solutions for three values of 
$\alpha=0.2,0.01,0.0005$ around nonrotating (Schwarzschild) and rotating 
(Kerr with $a=0.998$) black holes. 

\subsubsection{Schwarzschild black holes}

\begin{figure}
\begin{center}
\rotatebox{0}{\epsfig{file=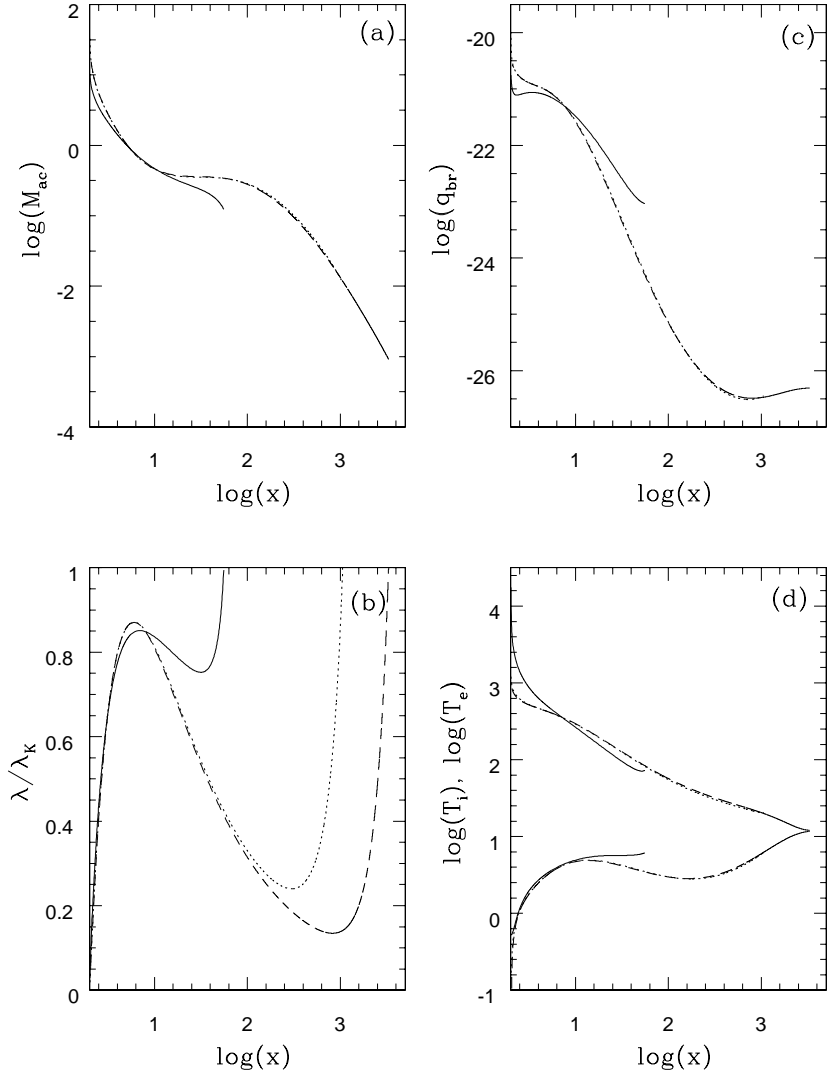,width=4.5in}}
\end{center}
\caption{ 
Same as Fig. \ref{fig1}, except for $M=10^7$, $\mdot=0.01$. 
Parameter sets are given in Table 2 in detail.
}\label{fig3}
\end{figure}

Table 2 lists the sets of input parameters.
Naturally a disk around a supermassive black hole will have much lower
density compared to that around a stellar mass black hole. Hence, the bremsstrahlung
radiation is expected to be less efficient leading to a radiatively inefficient flow,
particularly for a sub-Eddington accretion flow compared to a super-Eddington flow
around a stellar mass black hole.
However, the velocity profiles shown in Fig. \ref{fig3}a appear
similar to that around a stellar mass black hole, except with a stronger centrifugal barrier.
Due to dominance of gas (because of lack of supply of matter and then radiation), $\gamma$ is higher
and the disk appears puffed up and quasi-spherical further away compared to that around a stellar
mass black hole. Naturally $x_o$ recedes (Fig. \ref{fig3}b), 
where the hot flow is of two temperatures
unless $\alpha$ is very small rendering a high residence time of matter in the disk
which results in a relatively higher luminosity, given by Table 2.
However, as the disk around a supermassive black hole and 
the infall time scale of the matter therein is very large, the total luminosity is 
only two orders of magnitude lower, at the most, compared to that of stellar mass cases
described above, listed in Table 2.

\subsubsection{Kerr black holes}

\begin{figure}
\begin{center}
\rotatebox{0}{\epsfig{file=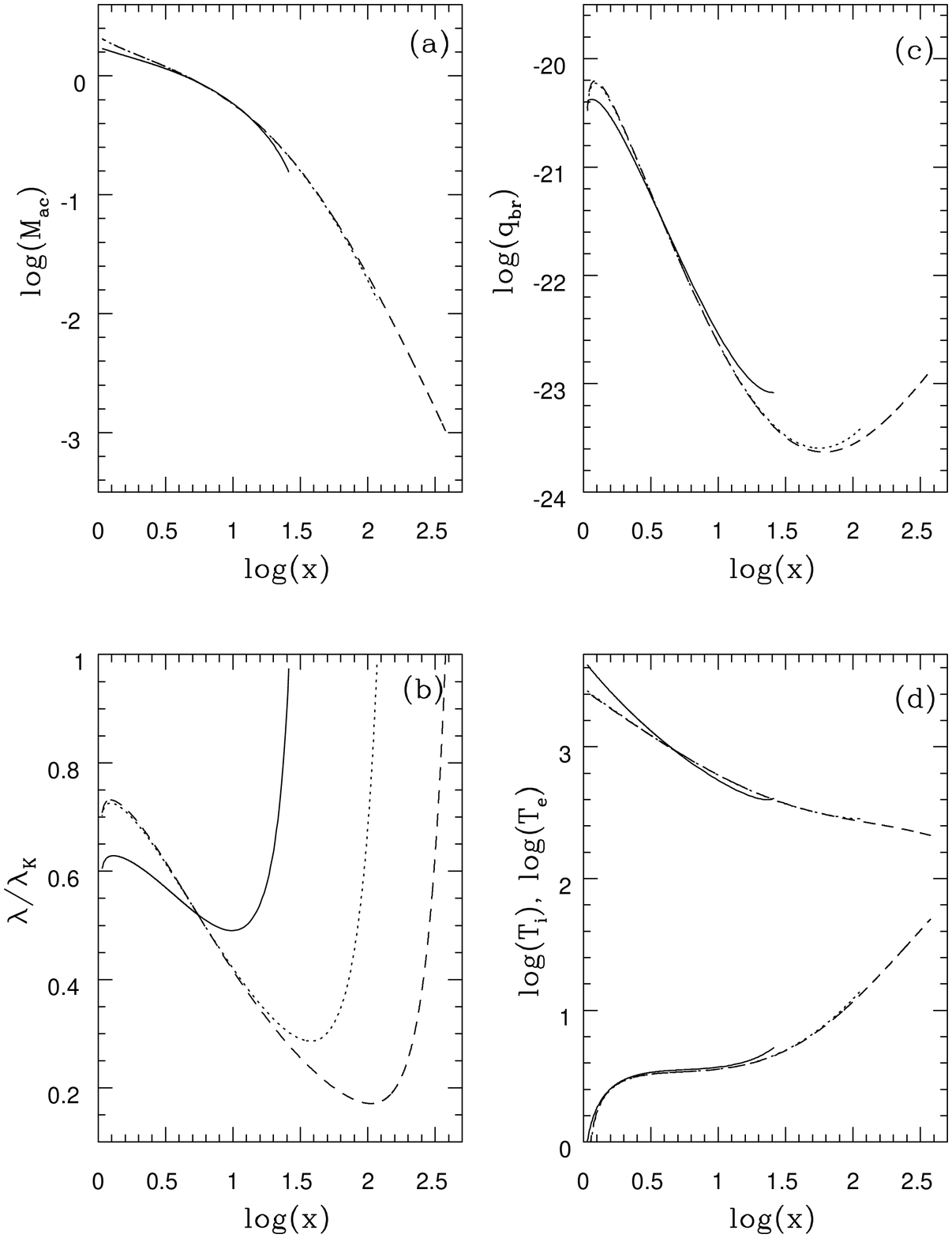,width=4.5in}}
\end{center}
\caption{ 
Same as Fig. \ref{fig3}, except for $a=0.998$.
}\label{fig4}
\end{figure}

The specific angular momentum of black hole is chosen to be $a=0.998$.
The basic hydrodynamic properties, particularly the velocity profiles, are similar 
to that around stellar mass black holes, except, like Schwarzschild cases, $x_o$
recedes, as explained above, shown in Fig. \ref{fig4}. 
As discussed in \S3.1.2, a higher $a$ corresponds to a smaller
$\lambda$ which in turn advances $x_o$ compared to that of the corresponding 
Schwarzschild case. As the outer edge of
sub-Keplerian flow is relatively closer to the event horizon for a rotating black hole,
due to stronger gravitational force ions are hotter and hence the flow is of two temperatures 
therein, even for $\alpha=0.0005$.
For similar reasons, as explained in \S3.1.2, the disk luminosity for a high $\alpha$
decreases compared to that of a static black hole, but increases for a very low $\alpha$
($=0.0005$).

\subsubsection{Under-luminous AGNs}

None of the above cases reveal the luminosity of an under-luminous source, e.g. Sgr~$A^*$.
Note that the computed luminosity strongly depends on the mass of the black hole and 
the supplied rate of accretion. If we choose $M=4.5\times 10^6$ 
(Ghez et al. 2008, Reid et al. 2008) and $\mdot\lsim 0.00001$ (Marrone et al. 2007), then the
luminosity tallies perfectly with that observed from Sgr~$A^*$, particularly
for a highly viscous flow (as is predicted by e.g. Narayan et al. 1995)
around a rotating black hole in a narrow range of the
Kerr parameter $a$. Table 3 shows how the luminosity
varies with the change of $\alpha$, $a$, as well as $f$. Note that we predict the possible
value of $a$ based on the extreme advection dominated cases (Narayan \& Yi 1995) 
when $f$ is strictly unity, as well as from
general advective paradigm with $0.5\lsim f\lsim 0.7$. Interestingly, for $f<1$, while $a\sim 0.2$
for a very highly viscous flow with $\alpha\sim 0.2$, it could be in a range $0.2-0.5$ 
when $\alpha\sim 0.05$. However, for advection dominated flows, $a\sim 0.2$.  
This argues Sgr~$A^*$ for a rotating black
hole with an intermediate spin $0.2\lsim a\lsim 0.5$.

\section{Comparison between flows around Schwarzschild and Kerr black holes}

From above discussions, we have already got an idea of 
basic differences between flows of same physical
parameters (e.g. $\alpha$, $\mdot$, $\gamma$) around static and rotating 
black holes of same mass. Here we present a one to one comparison for 
super-Eddington flows around stellar mass black holes. While, due to
general relativistic effects, a disk around a rotating black hole is
stretched further in, the transition zone advances quite a bit. Therefore,
the total size of sub-Keplerian disk decreases.

\begin{figure}
\begin{center}
\rotatebox{0}{\epsfig{file=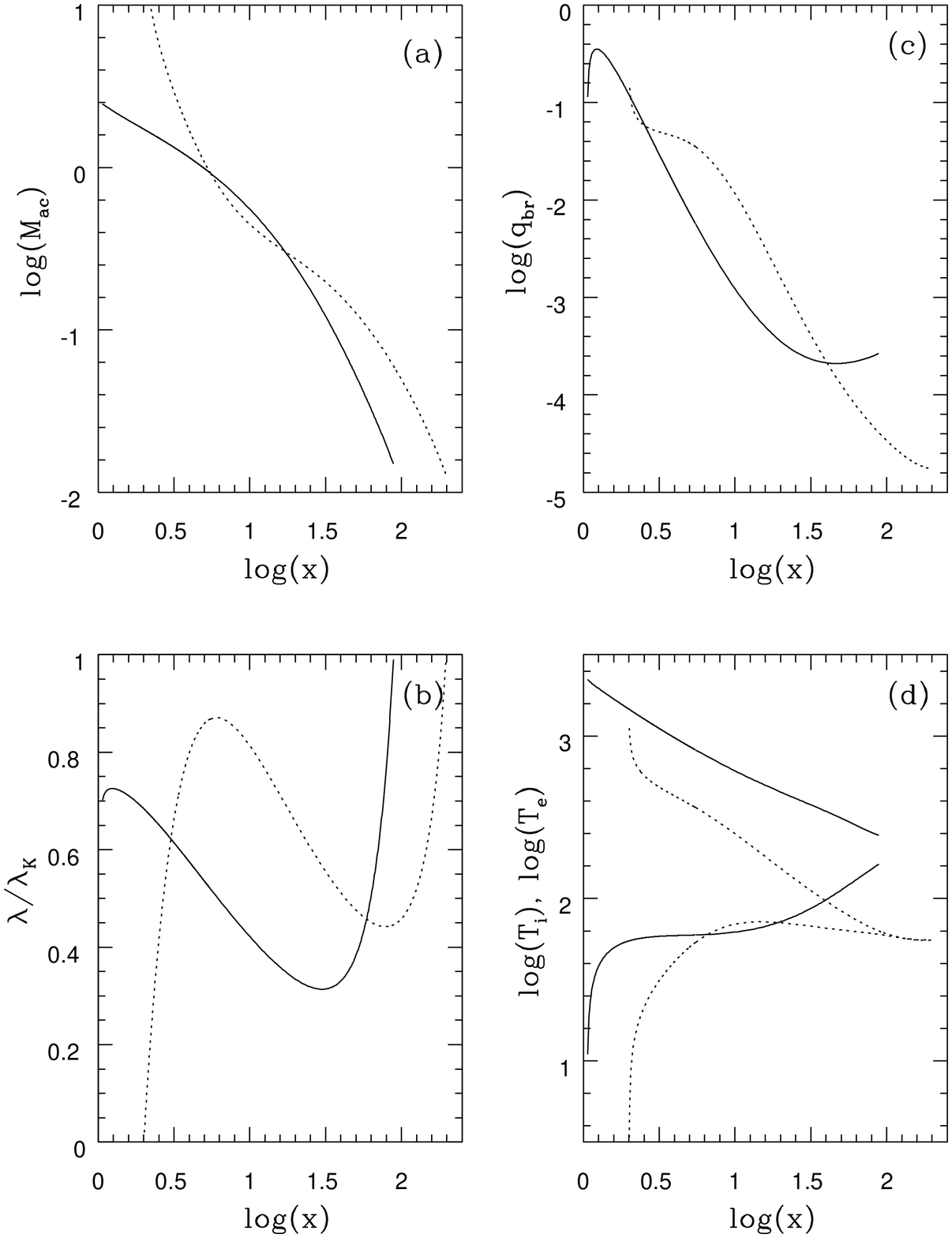,width=4.5in}}
\end{center}
\caption{ 
Variation of (a) Mach number, (b) ratio of disk angular momentum to corresponding
Keplerian angular momentum, (c) bremsstrahlung radiation in erg/sec, (d) electron
(lower set of curves) and ion (upper set of curves) temperatures in units of $10^9$K, as functions
of radial coordinate around a black hole of $M=10$ and $\mdot=10$. 
The solid, dotted curves correspond to $a=0.998,0$ respectively.
The respective parameter sets are given in Table 1 in detail.
}\label{fig5}
\end{figure}

From Fig. \ref{fig5}a it is clear that the velocity gradient of inner disk flow
around a rotating black hole is less stepper compared to that around 
a nonrotating black hole, e.g. in the vicinity of $x=2$, which is the event horizon 
for the latter case. Therefore, the radial velocity is lower in the former flow
at inner disk radii. However, the situation is opposite far away when
the sub-Keplerian flow ends at a smaller radius for a rotating black hole
and hence the disk is hotter and 
of two temperatures therein (Fig. \ref{fig5}d), unlike the Schwarzschild case.
Indeed the centrifugal barrier disappears for $a=0.998$. The radiative loss (Fig. \ref{fig5}c) also shows
opposite trends between flows around Kerr and Schwarzschild black holes, particularly
at the inner edge, as explained in \S3.1.2.

\section{Discussion and Summary}

We have modeled two temperature accretion flows around
black holes. We particularly focus on how the flow properties vary in a wide
range of $\alpha$, and between nonrotating and rotating black holes.
In this exploratory study, for simplicity, we consider only the bremsstrahlung 
radiation as a cooling mechanism. As the solutions for a low $\alpha$ ($=0.0005$) to 
a very high $\alpha$ ($=0.2$) reveal a significant variation of
properties/luminosities, in near future we plan to analyze more general solutions including 
all the possible cooling mechanisms (Sinha, Rajesh \& Mukhopadhyay 2009;
Rajesh \& Mukhopadhyay 2009).
 
In obtaining associated hydrodynamic solutions needed for computing 
the bremsstrahlung radiation and then the electron temperature profile,
we have strictly followed the earlier works (Chakrabarti\,\,1996; MG03). 
The temperature of the flow depends on the accretion rate. 
If the accretion rate is low and thus flow is radiatively 
inefficient (or less efficient), then the disk is hot. Such a hot flow is being
attempted to model since the work by Shapiro, Lightman \& Eardley (1976)
when it was assumed that $Q^+\sim Q^-$ locally and thus $f\rightarrow 0$. 
While the model successfully explained the observed hard X-rays from Cyg~X-1, it
turned out to be thermally unstable. Latter the hot ion torus model was proposed by
Rees et al. (1982) choosing $f<1$. In the
similar spirit Narayan \& Yi (1995) proposed the two 
temperature solution in the regime of $f\rightarrow 1$
when advection is very strong and then Yuan (2001) modified it
relaxing the constraint on the accretion rate. 
Mandal \& Chakrabarti (2005) proposed another two temperature disk
solution where the ion temperature could be as high as $\sim 10^{12}$K.
However, their particular emphasis was to understand the effect of shock to the 
cooling mechanisms. They do not carry out
a complete analysis of the flow. The present paper, to best of our knowledge,
describes for the first time the two temperature accretion flows around
{\it rotating} black holes in a wide range of $\alpha$.

In order to understand the implications of our solutions, we have chosen 
two extreme natural cases, namely observed ULX and under-luminous sources
which are respectively flows around stellar mass and supermassive black holes.
Interestingly the present model is able to explain the observed luminosities
for both the sources, as listed in Tables 1,2,3. 
It is revealed that at a very low mass accretion rate, $\mdot\lsim 0.00001$, around a rotating 
supermassive black hole of $M\gsim 10^6$, the luminosity comes out to be $L\gsim 10^{33}$ erg/sec, which
is indeed similar to the observed luminosity from an under-luminous 
source Sgr~$A^{*}$. Note that the computed $L$ increases (higher than that we observe from under-fed
sources) not only due to increase of $M$ and $\mdot$, 
but also for $a\rightarrow 0$. Therefore, we argue Sgr~$A^{*}$ for a spinning
black hole, presumably with a intermediate spin $a\lsim0.5$.
On the other hand, when we have considered a stellar mass black hole at a super-Eddington
mass accretion rate, $\mdot=10$, the model reveals
$L\sim 10^{41}$ erg/sec which is similar to the observed luminosity from
ULX sources. If the mass of the source is considered to be $M=10$, then
the flow must be highly viscous in order to attribute observed luminosity, 
according to the cases considered here.

Note that around a rotating black hole the flow angular momentum decreases and thus the 
radial velocity increases.
This in turn reduces the residence time of sub-Keplerian flow 
hindering cooling processes to complete. This flow with a high viscosity is then expected
to be low luminous compared to that around 
a static black hole. 
However, the luminosity also depends on the value of $\alpha$. Lowering $\alpha$ of the flow
increases residence time of matter in the disk which helps in cooling processes to
complete, switching over a low luminous flow to a high luminous one, when particularly 
the inner edge, which is radiatively most efficient in the disk, stretches in around a rotating black hole.
This feature may help in understanding the transient X-ray sources.

In all the cases, the ion and electron temperatures merge or
tend to merge at around transition radius, particularly for a high $\mdot$. This is because,
electrons are in thermal
equilibrium with ions and thus virial around the transition radius.
As the flow advances in the sub-Keplerian regime,
the ions become hotter and the corresponding temperature increases, 
rendering the ion-electron Coulomb collisions weaker.
Electrons, on the other hand, cool down via bremsstrahlung radiation
keeping the electron temperature roughly constant
upto very inner disk. This reveals the two temperature flow entirely.

Important point to note is that we have neither considered a very low
viscosity nor a very low accretion rate when nonthermal processes
of transferring energy from ions to electrons could be important
(Phinney 1981, Begelman \& Chiueh 1988). Instead, the viscous heating rate 
of ions is much higher
than the collective rate of nonthermal heating of electrons, unless
$\alpha$ is too small. Therefore, our assumption of coupling 
between ions and electrons due to the Coulomb scattering throughout 
is justified.
However, the process of transferring energy to electrons through
anisotropic pressure due to magnetorotational instability in collisionless 
plasma (Sharma et al. 2007) may be effective, particularly for the
low mass accretion rate, which we have neglected in the present work.

Now one should try to understand the radiation emitted
from the two temperature flows and to model the corresponding spectra.
This will be very useful in order to explain observed data.

\section*{Acknowledgments}
The authors would like to thank the referee for bringing in
attention to recent papers discussing mass and accretion rate of Sgr $A^*$.
This work is partly supported by a project, Grant No. SR/S2HEP12/2007, funded
by DST, India. One of the authors (SRR) acknowledges the Council for Scientific and
Industrial Research (CSIR; Government of India) for providing a research fellowship.

\clearpage

\noindent{ Table 1: Parameters and luminosity $L$ in erg/sec 
for accretion around stellar mass black holes,
when the subscript `c' indicates the quantity at the
sonic radius and $T_{ec}$ is expressed in units $10^9$K }

\begin{center}
\begin{tabular}{lllllllllllll}
\hline
\hline
$M$ & $\mdot$ & $a$ & $\gamma$ & $x_{c}$ & $\lambda_{c}$ & $\alpha$ & $f$ & $T_{ec}$ & $L$ \\
\hline
\hline
10 & 10 & 0     & 1.34  & 5.5 & 3.2 & 0.2 & 0.7 & 48 & $8.3\times 10^{41}$ \\
10 & 10 & 0 & 1.34  & 5.5 & 3.2 & 0.01 & 0.5 & 43.25 & $3\times 10^{42}$ \\
10 & 10 & 0 & 1.34  & 5.5 & 3.2 & 0.0005 & 0.3 & 43.1 &  $2.7\times 10^{43}$\\
\hline
10 & 10 & 0.998     & 1.34  & 5.5 & 1.5 & 0.2 & 0.7 & 50 & \,\,\,\,\,$10^{41}$\\
10 & 10 & 0.998 & 1.34  & 5.5 & 1.5 & 0.01 & 0.5 & 48 & $2.1\times 10^{42}$\\
10 & 10 & 0.998 & 1.34  & 5.5 & 1.5 & 0.0005 & 0.3 & 48 & $5.7\times 10^{43}$\\
\hline
\hline
\end{tabular}
\end{center}

\noindent{ Table 2: Parameters and luminosity $L$ in erg/sec 
for accretion around supermassive black holes,
when the subscript `c' indicates the quantity at the
sonic radius and $T_{ec}$ is expressed in units of $10^9$K }

\begin{center}
\begin{tabular}{lllllllllllll}
\hline
\hline
$M$ & $\mdot$ & $a$ & $\gamma$ & $x_{c}$ & $\lambda_{c}$ & $\alpha$ & $f$ & $T_{ec}$ & $L$  \\
\hline
\hline
$10^7$ & 0.01 & 0     & 1.5  & 5.5 & 3.1 & 0.2 & 0.7 & 2.92 & $4.2\times 10^{40}$\\
$10^7$ & 0.01 & 0     & 1.5  & 5.5 & 3.2 & 0.01 & 0.5 & 2.8 & $6.5\times 10^{40}$\\
$10^7$ & 0.01 & 0     & 1.5  & 5.5 & 3.2 & 0.0005 & 0.3 & 2.79 & $2\times 10^{42}$\\
\hline
$10^7$ & 0.01 & 0.998     & 1.5  & 5.5 & 1.5 & 0.2 & 0.7 & 2.85 & $2.6\times 10^{39}$\\
$10^7$ & 0.01 & 0.998     & 1.5  & 5.5 & 1.5 & 0.01 & 0.5 & 2.75 & $6.3\times 10^{40}$\\
$10^7$ & 0.01 & 0.998     & 1.5  & 5.5 & 1.5 & 0.0005 & 0.3 & 2.75 & $5.8\times 10^{42}$\\
\hline
\hline
\end{tabular}
\end{center}
\newpage
\noindent{ Table 3: Parameters and luminosity $L$ in erg/sec 
for accretion around Sgr $A^*$,
when the subscript `c' indicates the quantity at the
sonic radius and $T_{ec}$ is expressed in units of $10^9$K,
$M=4.5\times 10^6$, $\mdot\lsim 10^{-5}$ }

\begin{center}
\begin{tabular}{lllllllllllll}
\hline
\hline
$a$ & $\gamma$ & $x_{c}$ & $\lambda_{c}$ & $\alpha$ & $f$ & $T_{ec}$ & $L$  \\
\hline
\hline
 0     & 1.53  & 5.5 & 3 & 0.2 & 0.7 & 2.85 & $2.9\times 10^{33}$\\
 0     & 1.53  & 5.5 & 3.1 & 0.05 & 0.5 & 2.85 & $5.1\times 10^{33}$\\
\hline
 0.998     & 1.53  & 5.5 & 1.5 & 0.2 & 0.7 & 3.85 & $7.5\times 10^{32}$\\
 0.998     & 1.53  & 5.5 & 1.8 & 0.05 & 0.5 & 3.85 & $1.6\times 10^{33}$\\
\hline
 0.5     & 1.53  & 5.5 & 2.1 & 0.2 & 0.7 & 2.85 & $8\times 10^{32}$\\
 0.5     & 1.53  & 5.5 & 2.3 & 0.05 & 0.5 & 2.85 & $1.7\times 10^{33}$\\
\hline
 0.2     & 1.53  & 5.5 & 2.7 & 0.2 & 0.7 & 2.85 & $2\times 10^{33}$\\
 0.2     & 1.53  & 5.5 & 2.8 & 0.05 & 0.5 & 2.85 & $2.7\times 10^{33}$\\
\hline
\hline
 0     & 1.53  & 5.5 & 2.7 & 0.2 & 1 & 2.85 & $1.4\times 10^{33}$\\
 0     & 1.53  & 5.5 & 3 & 0.05 & 1 & 2.85 & $4.5\times 10^{33}$\\
\hline
 0.998     & 1.53  & 5.5 & 1.2 & 0.2 & 1 & 3.85 & $5.7\times 10^{32}$\\
 0.998     & 1.53  & 5.5 & 1.7 & 0.05 & 1 & 3.85 & $1.5\times 10^{33}$\\
\hline
 0.5     & 1.53  & 5.5 & 1.7 & 0.2 & 1 & 2.85 & $4.9\times 10^{32}$\\
 0.5     & 1.53  & 5.5 & 2.4 & 0.05 & 1 & 2.85 & $1.2\times 10^{33}$\\
\hline
 0.2     & 1.53  & 5.5 & 2.2 & 0.2 & 1 & 2.85 & $9.1\times 10^{32}$\\
 0.2     & 1.53  & 5.5 & 2.7 & 0.05 & 1 & 2.85 & $2.5\times 10^{33}$\\
\hline
\hline
\end{tabular}
\end{center}

\clearpage

\clearpage

\bigskip

\begin{thebibliography}{}

 \bibitem[]{921} Abramowicz, M. A., Czerny, B., Lasota, J. P., \& Szuszkiewicz, E.
1988, ApJ, 332, 646.
 \bibitem[]{922} Begelman, M. C. 1978, MNRAS, 184, 53.
 \bibitem[]{923} Begelman, M. C., \& Chiueh, T. 1988, ApJ, 332, 872.
 \bibitem[]{923} Begelman, M. C., King, A. R., \&  Pringle, J. E. 2006,
MNRAS, 370, 399.
\bibitem[]{921} Benson, A. J., \& Babul, A. 2009, MNRAS (to appear); arXiv0905.2378.

 \bibitem[]{924}{} Bisnovatyi-Kogan, G. S., \& Lovelace, R. V. E. 2000, ApJ, 529, 978.
 \bibitem[]{926}{} Chakrabarti, S. K. 1996, ApJ, 471, 237.
\bibitem[]{926}{} Chan, C., Psaltis, D., \& Ozel, F. 2005, ApJ, 628, 353.
\bibitem[]{926}{} Das, S., \& Chattopadhyay, I. 2008, NewA, 13, 549.
 \bibitem[]{948}{} Eggum, G. E., Coroniti, F. V., \& Katz, J. I. 1985,
ApJ, 298, 41.
 \bibitem[]{958}{} Eggum, G. E., Coroniti, F. V., \& Katz, J. I. 1988,
ApJ, 330, 142.
 \bibitem[]{968}{} Fabbiano, G. 2004, RMxAC, 20, 46.
 \bibitem[]{978}{} Fabrika, S. 2004, ASPRv, 12, 1.
 \bibitem[]{978}{} Ghez, A. M., et al. 2008, ApJ, 689, 1044.
 \bibitem[]{911}{} Ghosh, S., \& Mukhopadhyay, B. 2009, RAA, 9, 157.
 \bibitem[]{911}{} Ghosh, S., \& Mukhopadhyay, B., Krishan, V., \& Khan, M. 2010, 
NewA, 15, 83; arXiv:0906.0149.
 \bibitem[]{918}{} Gilfanov, M., Churazov, E., \& Sunyaev, R. 1997, LNP, 487, 45.
 \bibitem[]{912}{} Liang, E. P. T., \& Thompson, K. A. 1980, 240, 271.
\bibitem[]{912}{} Lipunov, V., \& Gorbovskoy, E. 2007, ApJ, 665, 97.
 \bibitem[]{914}{} Lovelace, R. V. E., Romanova, M. M., \&
Newman, W. I. 1994, ApJ, 437, 136.
 \bibitem[]{915}{} Mandal, S., \& Chakrabarti, S. K. 2005, A\&A, 434, 839.
 \bibitem[]{915}{} Marrone, D. P., Moran, J. M., Zhao, J.-H., \& Rao, R. 2007, ApJ, 654, 57.
 \bibitem[]{928}{} Matsumoto, R., Kato, S., Fukue, J., \& Okazaki, A. T. 1984,
PASJ, 36, 71.
 \bibitem[]{928}{} Melia, F., \& Misra, R. 1993, ApJ, 411, 797.
 \bibitem[]{928}{} Muchotrzeb, B., \& Paczynski, B. 1982, AcA, 32, 1.
 \bibitem[]{928}{} Mukhopadhyay, B. 2002, ApJ, 581, 427.
 \bibitem[]{928}{} Mukhopadhyay, B. 2003, ApJ, 586, 1268.
 \bibitem[]{911}{} Mukhopadhyay, B., \& Ghosh, S. 2003, MNRAS, 342, 274; MG03.
 \bibitem[]{914}{} Narayan, R., \& Yi, I. 1995, ApJ, 444, 231.
 \bibitem[]{914}{} Narayan, R., Yi, I., \& Mahadevan, R. 1995, Nature, 374, 623.
 \bibitem[]{921}{} Novikov, I. D., \& Thorne, K. S. 1973, in Black Holes, Les Houches 1972 (France), ed. B. \& C. DeWitt (New York: Gordon \& Breach), 343.
 \bibitem[]{928}{} Paczynsky, B., \& Wiita, P. J. 1980, A\&A, 88, 23.
 \bibitem[]{928}{} Phinney, E. S. 1981, in Plasma Astrophysics, ed. T. D. Guyenne \& G. Levy
(ESA SP-161), 337.
 \bibitem[]{911}{} Rajesh, S. R., \& Mukhopadhyay, B. 2009, MNRAS (to appear); arXiv:0910.4502.
 \bibitem[]{925}{} Rees, M. J., Begelman, M. C., Blandford, R. D., \& Phinney, E. S.
1982, Nature, 295, 17.
 \bibitem[]{928}{} Reid, M. J., Broderick, A. E., Loeb, A., Honma, M., \& 
Brunthaler, A. 2008, ApJ, 682, 1041.
 \bibitem[]{928}{} Shafee, R., Narayan, R., \& McClintock, J. E. 2008, ApJ, 676, 549.
 \bibitem[]{928}{} Shakura, N., \& Sunyaev, R. 1973, A\&A, 24, 337.
 \bibitem[]{928}{} Shapiro, S. L., Lightman, A. P., \& Eardley, D. M.	
1976, ApJ, 204, 187.
 \bibitem[]{928}{} Sharma, P., Quataert, E., Hammett, G. W., \& Stone, J. M. 2007,
ApJ, 667, 714.
 \bibitem[]{928}{} Sinha, M., Rajesh, S. R., \& Mukhopadhyay, B. 2009, RAA (to appear); arXiv:0910.4818.
 \bibitem[]{928}{} Stuchl\'ik, Z., \& Kov\'ar, J. 2008, IJMPD, 17, 2089.
 \bibitem[]{928}{} Yuan, F. 2001, MNRAS, 324, 119.


\end{thebibliography}
\end{document}